\documentstyle[prl,twocolumn,aps]{revtex}
\input psfig.sty
\input epsf.sty
\begin{document}
\draft
\title{Evidence for Suppression of Superconductivity by Spin Imbalance
in Co-Al-Co single electron transistors }
\author{C. D. Chen$^1$\cite{email}, Watson Kuo$^{1,2}$, D. S. Chung$^1$,
J. H. Shyu$^{1,3}$, C. S. Wu$^1$}
\address{ $^1$Institute of Physics, Academia Sinica,
Nankang, 115, Taipei, Taiwan, ROC \\ $^2$Department of Physics, National
Tsing-Hua University, Hsin-Chu, 300, Taiwan, ROC \\ $^3$Department of
Physics, National Taiwan University, Taipei, Taiwan, ROC}
\date{March 25, 2001}
\maketitle
\begin{abstract}
Spin imbalance is predicted to lead to suppression of superconductivity.
We report phenomena manifesting this effect under spin-polarized
quasiparticle currents in ferromagnet-superconductor-ferromagnet single
electron transistors. The measured superconducting gap as a function of
magnetic field reveals a dramatic decrease when the magnetizations of the
two leads are in opposite orientations. The effect of suppression
increases with increasing voltage but decreases at elevated temperatures.
The possible explanations for these dependences are given. This method may
render it applicable to control superconductivity at low temperatures and
low fields.
\end{abstract}


The interplay between superconductivity and magnetism has been a topic of
interest for many years. Of particular interest are recent experiments on
ferromagnet/superconductor (FM/SC) junctions with high $T_C$\cite{1} in
which a decrease of the supercurrent by nonequilibrium spin density was
demonstrated. Furthermore, theoretical studies\cite{2,3} also indicate
that spin imbalance in a superconductor can lead to suppression of
superconductivity. In a double tunnel junction containing a normal metal
or a superconductor sandwiched between two ferromagnets, both injection of
polarized current and spin accumulation are possible, and thus provide an
ideal test ground for the theory. When the magnetic moments of the two
ferromagnetic leads are in opposite orientations (referred to as the
antiferromagnetic (AF) alignment), the difference in the number of
majority and minority spins in the central electrode brings in a chemical
potential difference. In a normal metal, this potential difference is
superconductors $\delta\mu=PeV/2$, where $P$ is the polarization of the
ferromagnetic leads and $V$ is the voltage across the sample. In a
superconductor, this difference gives rise to pair breaking, same as the
Zeeman effect does to superconductivity in the paramagnetic limit. In this
paper, we report the first direct observation of the superconducting gap
suppression using Co/Al/Co (FM/SC/FM) double tunnel junctions. The effect
can be turned on and off by manipulating the mutual orientations of the
magnetic moment of the two Co leads. We further discuss some relevant time
scales, including the spin relaxation time and the tunneling time, which
set the criterion for spin accumulation in the superconductor.

The inset of Fig.\ 1 shows an SEM picture of a measured sample and its
biasing circuit. The samples were fabricated by standard electron-beam
lithography techniques and by the shadow evaporation method. A thin native
${\rm Al_2O_3}$ layer between the Al island and the Co electrodes acted as
a tunnel barrier. Aluminum is a good candidate to serve as the
superconductor for this purpose, not just for its long spin
lifetime\cite{4} allowing the full range of spin effects to be studied,
but also for its high quality native ${\rm Al_2O_3}$ barrier which was
shown to have no spin-flip tunneling processes\cite{5}. Our electron-gun
deposited Al islands have a superconducting transition temperature of
about 2K. The samples were measured using a dilution refrigerator and the
magnetic field was applied along the Co leads, {\it i.e.} parallel to the
long edge of the Al island. Because of the small thickness ($\approx$25nm)
of the Al island and of the alignment with the magnetic field, the
critical magnetic field $H_{C\parallel}$ for the Al islands was 21kGs.
Experimentally, we found that the coercivity field for the two Co-leads
was smaller than 2kGs, well below $H_{C\parallel}$, and thus has
negligible influence on the superconductivity of the Al island. A
necessary condition for the imbalance is that the spin relaxation time ts
is longer than the time tt between two successive tunneling events. A
rough estimate of $\tau_t$ is $e/I$\cite{6}, which, as will be shown
below, is about 0.1$\sim$1 ns. This is shorter than the reported spin
relaxation time $\tau_s$ of about 10ns in Al films\cite{4}, making the
spin imbalance possible.

The current-voltage ($IV$) characteristics are similar to those for a
normal metal-superconductor-normal metal (N/SC/N) double junction
structure. Because the tunnel junction is small, the charging energy $E_C$
associated with it is large\cite{7}. The current-voltage characteristics
measured at low temperatures displayed, in addition to the superconducting
gap $2{\Delta_0}/e$, a pronounced Coulomb gap $2E_C/e$. This Coulomb gap
can be suppressed to a minimum by tuning the voltage applied on the gate
electrode, which is coupled to the central island via a capacitor $C_g$.
From a plot of the zero field $IV$ characteristics as a function of gate
voltages ({\it i.e.} the ``Coulomb Blockade parallelogram"\cite{8}), we
estimated a superconducting gap $\Delta_0$ of about 250 $\mu e{\rm V}$,
and a charging energy $E_C$ of 100$\sim$150 $\mu e{\rm V}$.

Throughout this study, the samples were symmetrically current-biased with
respect to the ground. We applied a parallel magnetic field to decrease
the size of the superconducting gap, and monitored this decrease by
measuring the voltages. As shown in Fig.\ 1, when the field was ramped
from +25kGs to -25kGs, we found a sharp voltage drop at a field of around
-1.5kGs; When the field was reversed and ramped up from -25kGs to around
+1.5kGs, then the voltage dropped again, yielding a symmetric $V(H)$
pattern with respect to zero field. Because of the current-bias circuitry,
the change in the current was less than 1\% regardless of the large
voltage change. This hysteresis was reproducible, and similar results were
found for other samples. This behavior is very similar to the hysteresis
seen in tunneling magnetoresistance (TMR) of a
ferromagnet-insulator-ferromagnet single tunnel junction (see for example
Ref. \cite{9}). By token of this similarity it is reasonable to argue that
a field of $|H|=25$kGs is sufficiently large to saturate the magnetization
of the two leads and to make them parallel to the applied fields, whereas
at small fields ($H=\pm 1.5$kGs for the sample of Fig.\ 1) they are in AF
alignment. The presence of the two dips in voltage in the AF alignment
thus demonstrates the suppression of superconductivity of the Al island,
due to spin imbalance.

Figure 1 also shows $V(H)$ curves taken at various bias currents. By
taking the $V(H)$ curves at various bias currents one can recover the $IV$
characteristics, as displayed in Fig.\ 2, for three different cases: (I)
When the island is in the superconducting state with no applied magnetic
field. (II) When the island is in the normal state and the two leads are
in the ferromagnetic alignment. (III) When the island is in the
superconducting state and the two leads are in the AF alignment. For the
case (I), the obtained $IV$ characteristic can be fitted by a simulation
\cite{10} of a N/SC/N single electron transistor at gate voltage $V_g$
corresponding to the minimum Coulomb blockade threshold ({\it i.e.}
$V_g=e/2C_g$). From this fitting, one can obtain a good estimate of the
sample parameters, including the capacitances and the resistances of the
two tunnel junctions and, most importantly, the zero-field superconducting
gap $\Delta_0$ of the Al island. The fitted $\Delta_0$ value is about 260
$\mu e{\rm V}$, agreeing well with the gap estimate from the plot of
``Coulomb Blockade parallelogram". The fitted junction parameters can then
be checked by a fitting to the $IV$ characteristics obtained for the case
(II), that is, for an all-normal-metal (N/N/N) transistor. The parameters
for this particular sample are as follows (with subscripts $S$ and $D$
stand respectively for source and drain): $R_S=R_D$=237kW, $C_S$=360aF,
$C_D$=300aF, $C_g$=0.47aF, and $V_g$=160mV. A temperature of 250mK is used
for this fitting; this is reasonable considering the heating due to eddy
current induced by the sweeping field. At a voltage interval, the
calculated current for the case (II) is smaller than the measured ones;
this may be attributed to tunneling processes that are not included in the
simulation. For example, it is known that for $V_g=e/2C_g$, there is a
knee at $V=2E_C/e=0.24$mV, below which sequential tunneling process
suppresses the differential conductance by a factor of two. However, in
reality this feature may be smeared probably due to Zener tunneling to the
upper band.\cite{11}

With the knowledge of the sample parameters one can calculate the $IV$
characteristics at $V_g=e/2C_g$ for various superconducting gaps
$\Delta_{\rm A}$. Note that in the AF alignment, the effective drain
tunneling resistance $R_{D{\rm A}}$ increases by a factor of
$(1+P^2)/(1-P^2)$. For the case of the Co electrodes the reported $P$
value is about 0.4.\cite{12} Taking into account this correction, we plot
the calculated curves together with the $IV$ curve for the cases
(I)$\sim$(III) in Fig.\ 2. The cross points between the $IV$ curve for the
case (III) and the calculated curves for various superconducting gaps
$\Delta_{\rm A}$ in the AF alignment give a relationship between the
voltage $V$ and the corresponding $\Delta_{\rm A}$, as depicted in Fig.\
3(a). Theoretically the dependence of $\Delta_{\rm A}$ on $V$ is
determined by combining the $\Delta_{\rm A}(\delta\mu)$ and
$\delta\mu(eV)$ functions, and self-consistent calculating the values of
$\delta\mu$, $eV$ and $\Delta_{\rm A}$.\cite{2} The calculated
$\Delta_{\rm A}/\Delta_0(V)$ curves are shown in Fig.\ 3(a). While the
theory is based on the assumption that majority spins can accumulate in
the island, this imbalance of spins is a nonequilibrium process. In view
of this, our analysis, which assumes a $R_{D{\rm A}}$ using a $P$ value of
the ferromagnetic leads, could be an oversimplified approach.
Nevertheless, this analysis should give a quantitative picture of the
$\Delta_{\rm A}/\Delta_0(V)$ dependence.

Our analysis (depicted in Fig.\ 3(a)) with $P$=0.4 shows that at
$V<2\Delta_0/e$ the ratio $\Delta_{\rm A}/\Delta_0(V)$ increases with
voltage and a maximum value appears at $V_{\rm peak}\approx
2.4\Delta_0/e$. To see the effect of the $P$ value on the $\Delta_{\rm
A}/\Delta_0(V)$ dependence, we further test a $P$ value of 0.6, which
corresponds to a higher $R_{D{\rm A}}$. As shown in Fig.\ 3(a), because of
this increase in resistance, the $V_{\rm peak}$ value shifts toward
$2\Delta_0/e$. Furthermore, for both $P$ values at $V$ below $V_{\rm
peak}$ we found suppression of $\Delta_{\rm A}$ to about $\Delta_0(V)$ and
a lowering of $\Delta_{\rm A}$ with decreasing $V$. This behavior is even
pronounced for sample \#2. Ideally, at low temperatures and low voltages,
absence of quasiparticle tunneling should lead to a $\Delta_{\rm
A}/\Delta_0(V)$ value close to 1. Various origins can give rise to excess
current, which are not taken into account in our simulation. This excess
current, in our analysis, can give a lower $\Delta_{\rm A}$ value than
what is expected. One possible source for such excess current is the
co-tunneling process.\cite{8}  In this process, the excess current
increases at voltages close to $\Delta_{\rm A}/e$, and has stronger
influence for $IV$ curves with smaller $\Delta_{\rm A}$ values, and is
thus responsible for the decrease of the $\Delta_{\rm A}/\Delta_0$ value
at lower voltages. On the other hand, for $V>V_{\rm peak}$ , the
calculated $\Delta_{\rm A}/\Delta_0$ value is smaller than the
experimental ones. This discrepancy probably arises from a too long energy
relaxation time in our small Al island, which can smear the effect of spin
imbalance.

We have also measured the $V(H)$ curves (as those in Fig.\ 1) at various
temperatures. By using the same analysis technique as above, we traced out
separately the $\Delta_0(T)$ and $\Delta_{\rm A}(T)$ dependences, and
plotted the ratio $\Delta_{\rm A}/\Delta_0$ as a function of temperature.
As shown in Fig.\ 3(b), for $T<0.6$K this ratio increases with
temperature. In this temperature range, both the energy relaxation time
and the spin relaxation time do not change much with temperature.  Our
result, shown in Fig. 3b, may be explained by the fast relaxation of
domains in the Co leads, as we discuss below.

In the TMR experiments, the two ferromagnetic electrodes are usually made
of two different materials with different coercivities. In our case, the
AF alignments may be caused by random domain distributions in the two Co
leads\cite{13,14}. The junction area is about 80nm$\times$80nm, which is
close to the reported domain size of a thin Co wire\cite{14}. Therefore,
the measured hysteresis shown in Fig. 1 is simply a sign of magnetization
reversal in probably a single domain. The magnetic field was swept at a
rate of 2.5kGs/min; this is a crucial condition for the observed strong
suppression of the superconducting gap. As shown in Fig.\ 4(a), a smaller
sweep rate results in a smaller suppression effect. This is a signature of
a lesser spin imbalance and indicates imperfect AF alignment of the two
leads. We note that this sweep rate dependence is completely reproducible,
and that similar results were found for several other samples. The fact
that the amount of spin imbalance is sweep rate dependent indicates that
the magnetization reversal is a nonequilibrium process. All magnetic
systems that exhibit hysteresis are expected to relax slowly toward the
free energy minimum. Indeed, when the field is swept from +25kGs to
-1.5kGs (at which $\Delta_{\rm A}$ is at a minimum) at a rate of 2.5kG/min
and pause, then the value of $\Delta_{\rm A}$ decrease with time.
Moreover, as shown in Fig.\ 4(b), the rate of this decrease is higher at
higher temperatures. The relaxation time follows a simple Arrhenius
dependence on temperature\cite{14}, indicating a thermal activated
relaxation of magnetization in individual domains. The activation energy
is about 140K. This relaxation can account for the lessening suppression
of $\Delta_{\rm A}$ at high temperatures.

In conclusion, the superconducting gap of a small Al island, when
incorporated in a single electron transistor structure in which the two
electrodes were made of Co, was found to be strongly suppressed by
polarized quasiparticle injection from the ferromagnetic leads if the two
leads were in AF alignment. The effect appears to be caused by pair
breaking associated with spin-imbalance in the superconducting island. At
elevated temperatures, faster relaxation of magnetization of the two
ferromagnetic leads yields a diminution of spin imbalance. This experiment
provides a new method of controlling superconductivity.

Fruitful discussions with S. K. Yip, G. Y. Guo, C. S. Chu, Y. S. Wu, J. J.
Lin and S. F. Lee are gratefully acknowledged. We have made use of
facilities at the National Nano Device Laboratories. This research was
partly funded by the Nation Science Council No. 89-2112-M-001-033.

\begin{figure}[tbp]
\caption{Measured voltages as a function of applied magnetic field for
several selected bias currents: from bottom, 0.05, 0.20, 0.35, 0.50, 0.60
and 0.75 nA. The inset shows an SEM image of one sample. The white object
in the center is the Al island, and the scale bar at the bottom is 100nm.}
\label{fig1}
\end{figure}

\begin{figure}
\caption{Bold curves: the $IV$ characteristics as recovered from Fig.\ 1
using the voltages and the corresponding bias currents at $B=\pm 25$kGs
(leftmost curve), the maximum voltage at $B\approx 0$ (rightmost curve),
and the voltages at the dips at $B\approx +1.5$kGs (middle curve). The two
solid curves are calculated respectively for the superconducting and
normal states, and the dashed curves are calculated for various
$\Delta_{\rm A}$ values (shown in the figure are 220,200,180$\mu e{\rm
V}$), as described in text. The insert shows measured large-scale $IV$
curves at $B=0$ and $B=25$kGs and the calculated $IV$ curves, which
confirms the obtained sample parameters. }\label{fig2}
\end{figure}

\begin{figure}
\caption{(a) $\Delta_{\rm A}/\Delta_0$ as a function of voltage, as
obtained from Fig.\ 2. Calculated curves for $T=$250mK are also plotted
for comparison.  (b) The ratio $\Delta_{\rm A}(T)/\Delta_0(T)$ as a
function of temperature for sample \#2. The suppression (and thus the
imbalance) decreases with temperature due to a decrease in lead's
magnetization.} \label{fig3}
\end{figure}

\begin{figure}[tbp]
\caption{(a) The ratio $\Delta_{\rm A}/\Delta_0$ as a function of the
magnetic field sweep rate. The value of $\Delta_{\rm A}$ decreases with
increasing sweep rate and saturates at rates $>2.0$kGs/min. (b) The
relaxation time of lead magnetization as a function of inverse temperature
which shows a thermally activated relaxation behavior.} \label{fig4}
\end{figure}


\begin{references}
\bibitem{6}From the continuity equations for electric charges and
spins one can see that in the steady state the spin imbalance
$(N_\uparrow-N_\downarrow)$ is related to the tunneling rates for, say, up
spins $\Gamma_{S\uparrow}$ and $\Gamma_{D\uparrow}$ on the source and
drain junctions respectively as
$(N_\uparrow-N_\downarrow)/\tau_s=\Gamma_{S\uparrow}(\delta\mu+V/2)-
\Gamma_{D\uparrow}(\delta\mu-V/2)$. In our experiments, the transistors
were operated in the sequential tunneling regime (only one electron can
tunnel at a time), and hence the tunneling rate is on the order of $e/I$,
which is to be compared with $1/\tau_s$
\end{references}
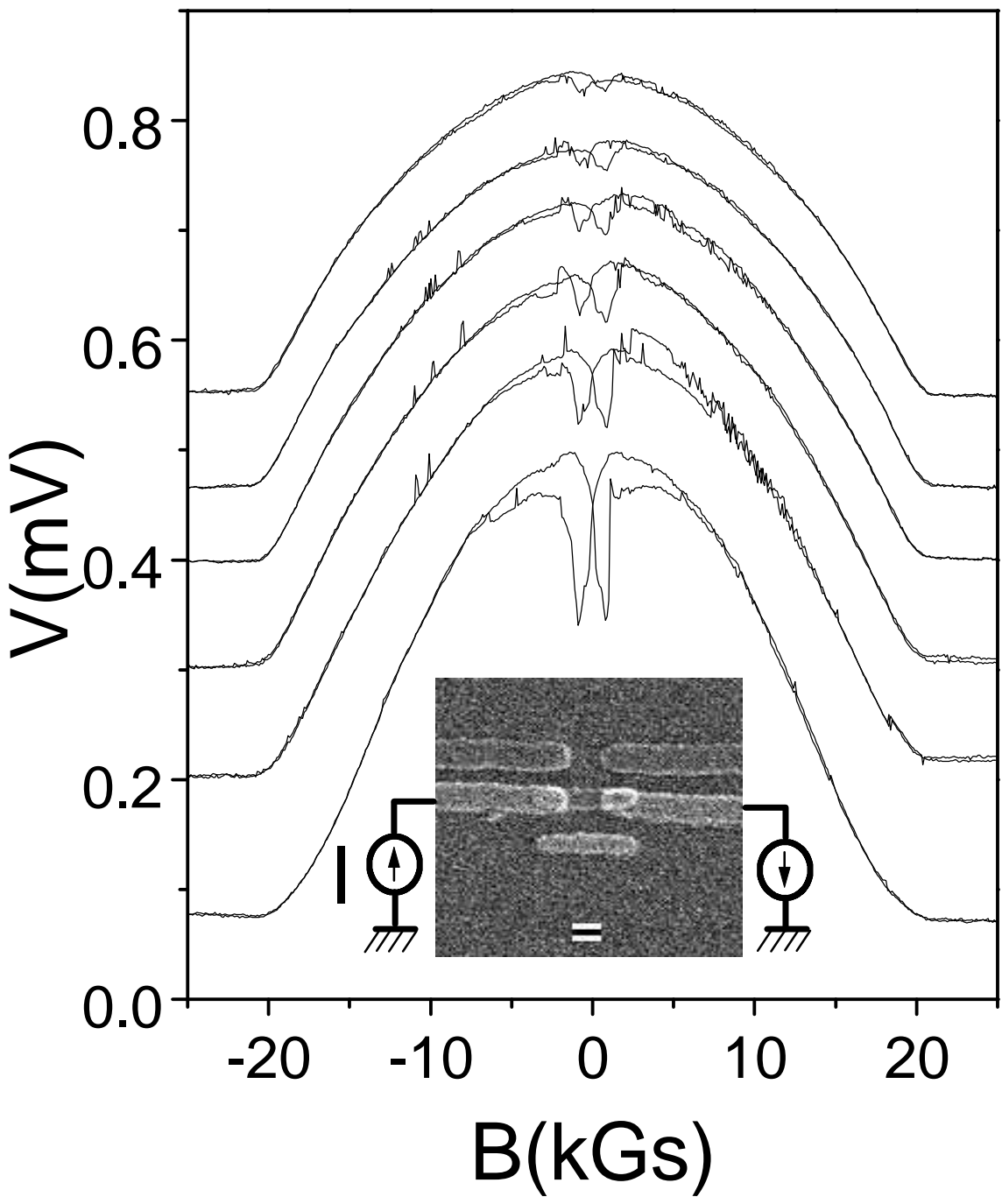
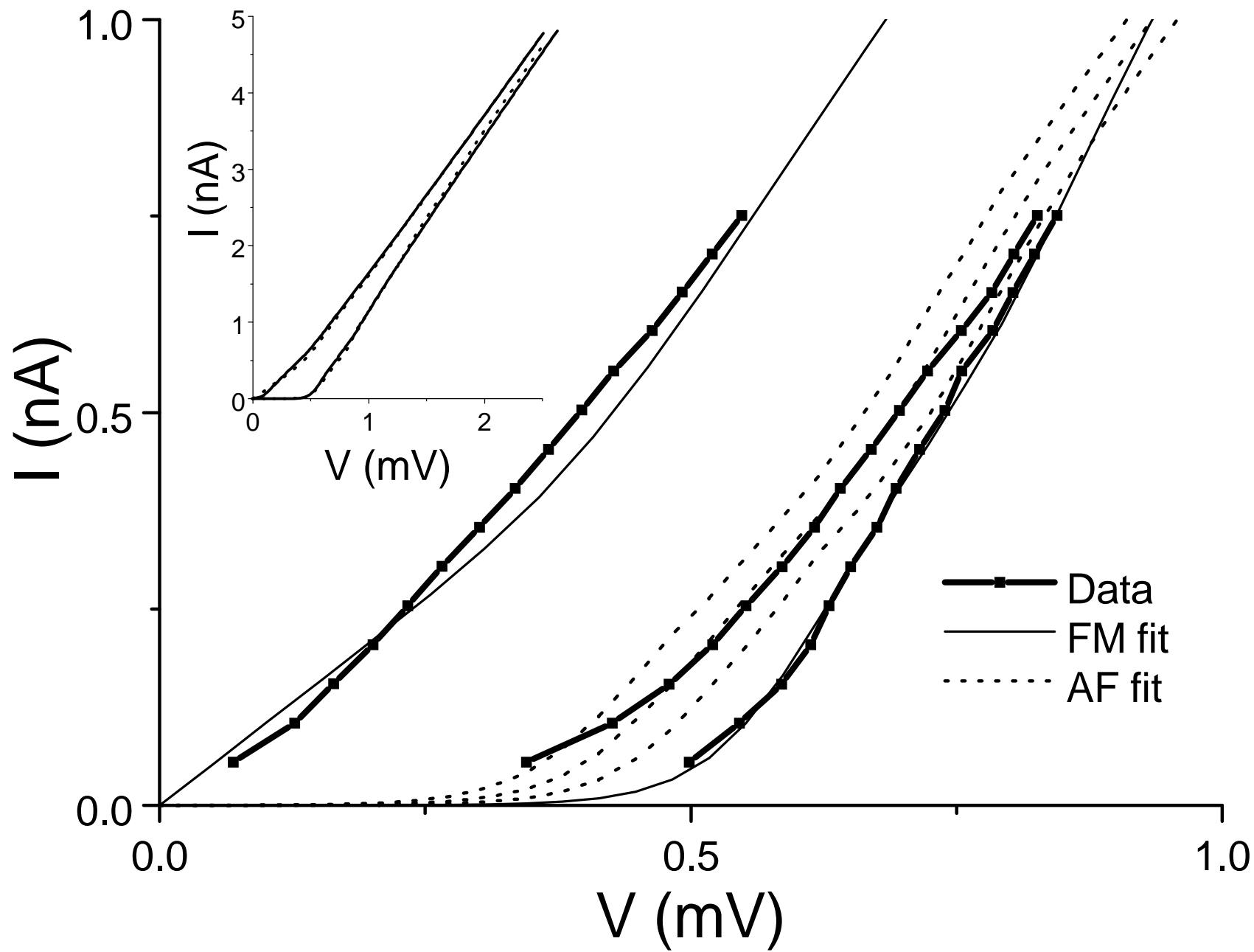
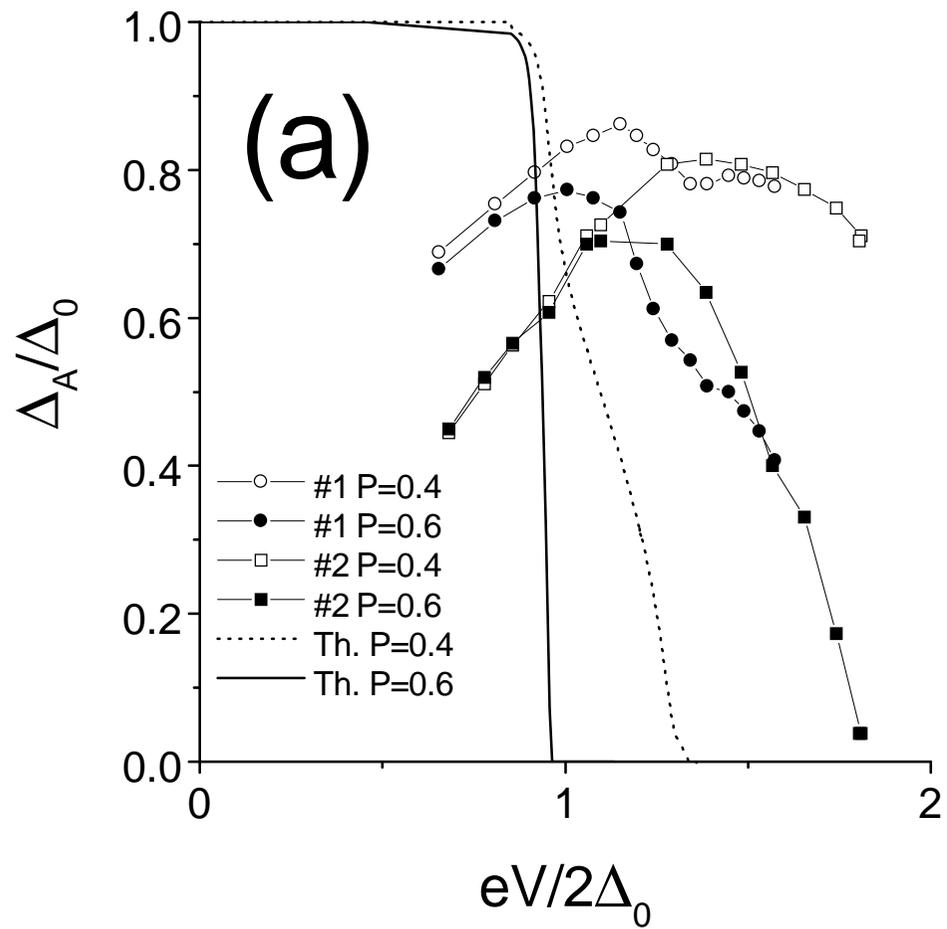
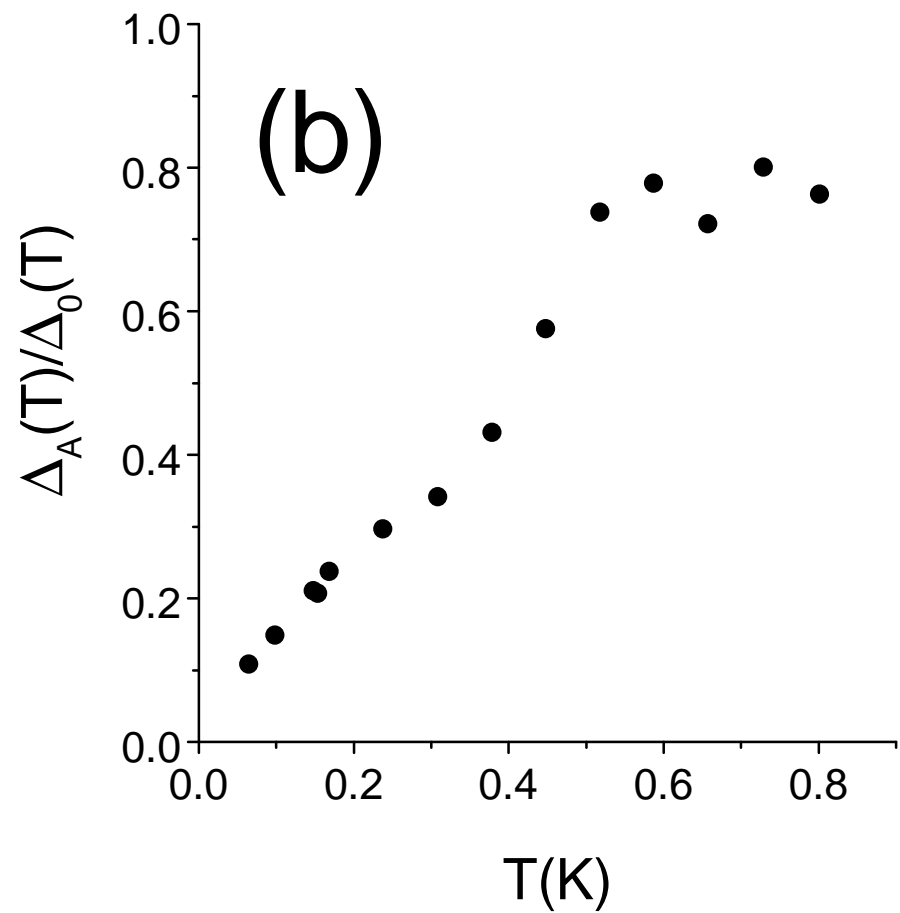
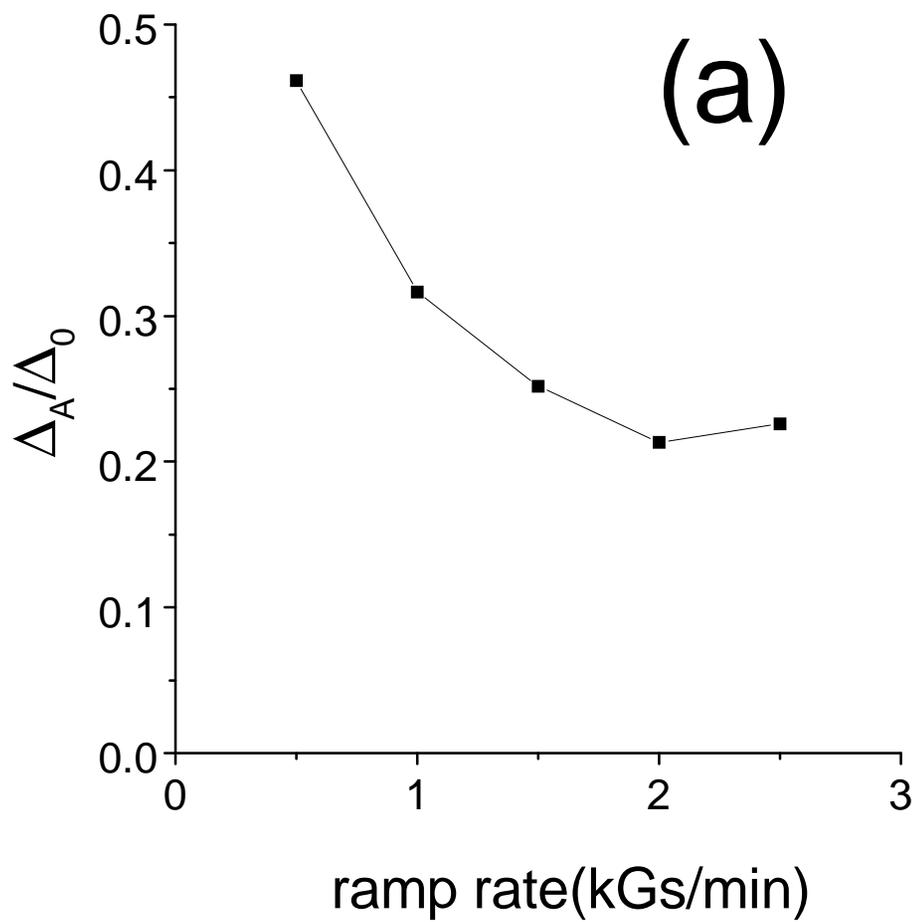
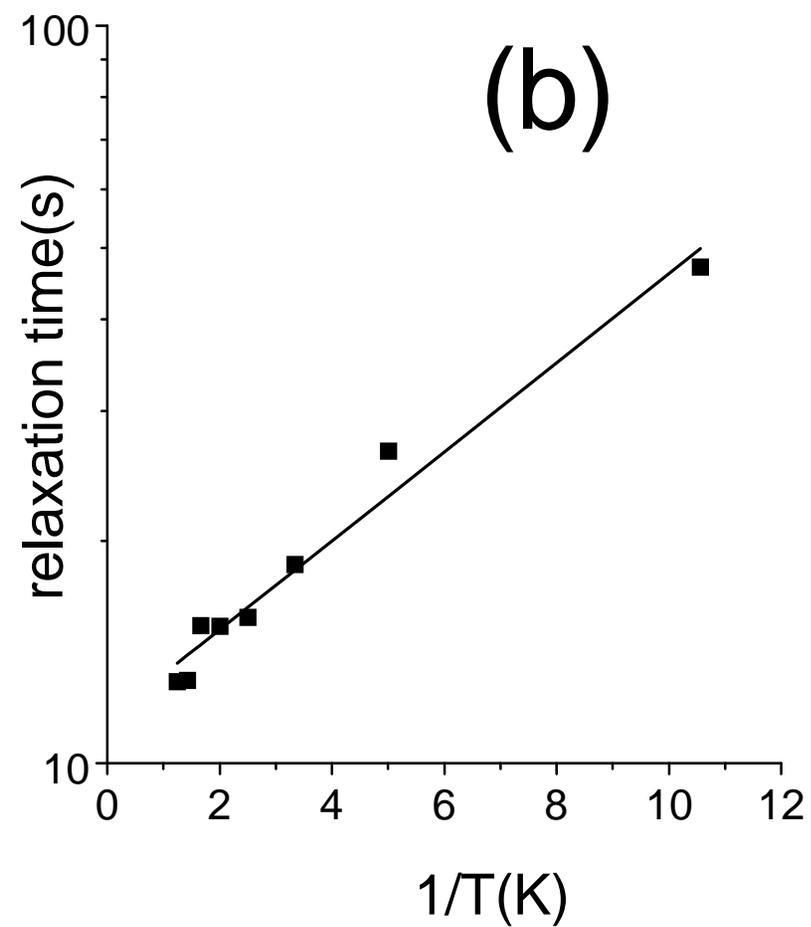
\end{document}